\documentclass[12pt]{article}
\usepackage{epsfig}
\usepackage{amsmath}
\usepackage{amssymb}
\usepackage{cite}
\usepackage{color}
\usepackage{graphicx}

\numberwithin{equation}{section}
\numberwithin{table}{section}

\newcommand{\be}{\begin{equation}}
\newcommand{\ee}{\end{equation}}

\DeclareMathOperator*{\nij}{\text{Nij}}

\begin{document}

\begin{titlepage}
\begin{flushright}
hep-th/0612100\\
MAD-TH-06-15
\end{flushright}

\vfil

\begin{center}

{\Large \bf NS-NS fluxes in Hitchin's generalized geometry}

\vfil

{ Ian T. Ellwood}

\vspace{.6in}

{ Department of Physics \\
University of Wisconsin-Madison\\ 
Madison, WI 53706, USA\\
E-mail: {\tt iellwood@physics.wisc.edu}}

\vfil

\end{center}


\begin{abstract}
\noindent 
The standard notion of NS-NS 3-form flux is lifted to Hitchin's
generalized geometry.  This generalized flux is given in terms of an
integral of a modified Nijenhuis operator over a generalized 3-cycle.
Explicitly evaluating the generalized flux in a number of familiar
examples, we show that it can compute three-form flux, geometric flux
and non-geometric $Q$-flux.  Finally, a generalized connection that
acts on generalized vectors is described and we show how the flux
arises from it.
\end{abstract}


\vspace{0.5in}

\end{titlepage}

\section{Introduction} \label{sect:introduction}

Generalized (complex) geometry, developed by Hitchin
\cite{Hitchin:2004ut,Hitchin:2005cv,Hitchin:2005in} and Gualtieri
\cite{Gualtieri:2003dx}, has emerged as a useful framework for
describing new string compactifications.  It naturally includes a
large class of vacua known as generalized Calabi-Yau manifolds
\cite{Grana:2004bg,Grana:2004sv,Grana:2005sn,Grana:2005ny,Shelton:2005cf,Shelton:2006fd,Benmachiche:2006df},
and also gives a more elegant description of so-called non-geometric
spaces or T-folds
\cite{Hellerman:2002ax,Dabholkar:2002sy,Kachru:2002sk,
Hull:2006va,Hull:2006tp,Hull:2006qs,Grange:2006es}.

A strength of this formalism is that it is naturally
covariant under T-duality provided one dualizes along a
$U(1)$-isometry direction.  While the action of T-duality on the sugra
fields, given by the Buscher rules, is very complex
\cite{Buscher:1987sk,Buscher:1987qj, Bergshoeff:1995as}, the corresponding
transformations in generalized geometry are quite simple.

One of the interesting features of generalized geometry is that the
metric and $B$-field are no longer considered the fundamental objects.
Rather, it is only the combination of $g$ and $B$ \cite{Duff:1989tf},
\begin{equation}\label{bigMetric}
  \mathsf{G}  =
\left( 
   \begin{matrix} 
     -g^{-1} B     & g^{-1} \\
     g - Bg^{-1} B & B g^{-1} 
   \end{matrix} 
  \right),
\end{equation}
that enters into the formalism.  This unification of $B$ and $g$ is
very natural in a T-duality covariant formalism since $B$ and $g$ mix
under T-duality.

However, having grouped $g$ and $B$ together into a single generalized
object, $\mathsf{G}$, we are left with a puzzle: What is the generalized
version of NS-NS 3-form flux, $H = dB$?  In particular, we would
like to find a generalized analogue of 
\begin{equation} \label{basicFluxIntegral}
  \int_\Sigma H \ ,
\end{equation}
where $\Sigma$ is a 3-cycle and $H = dB$ is the NS-NS 3-form flux.
Since generalized geometry is covariant under T-duality, the
generalized version of (\ref{basicFluxIntegral}) should also capture
its various T-duals.  Under T-duality, 3-form flux is mapped to
so-called geometric flux, which is given by the first Chern-class of a
circle bundle
\cite{Alvarez:1993qi,Gurrieri:2002wz,Kachru:2002sk,Bouwknegt:2003vb}.
Applying T-duality once more, the flux becomes the somewhat obscure
non-geometric flux
\cite{Kachru:2002sk,Grange:2006es,Lowe:2003qy,Mathai:2004qc,Bouwknegt:2004ap,Hull:2004in,Dabholkar:2005ve,Hull:2006va,Hull:2006qs,Benmachiche:2006df,Ellwood:2006my}
or $Q$-flux in the parlance of \cite{Shelton:2005cf,Shelton:2006fd}.

The purpose of this paper is to argue that the generalized analogue of
$H$, which we denote $\mathsf{H}$, is a slight modification of the
Nijenhuis operator given in \cite{Gualtieri:2003dx}.  Just as $H$,
being a three-form, can be defined by its action on vectors,
$H(V_1,V_2,V_3)$, we define $\mathsf{H}$ by its action on {\em
generalized} vectors $\mathsf{V} \in T\oplus T^*$,
\begin{equation} \label{GeneralizedHFlux}
  \mathsf{H}(\mathsf{V}_1, \mathsf{V}_2,\mathsf{V}_3) = -\mathsf{Nij}
 ( \widetilde{\mathsf{V}}_1,
   \widetilde{\mathsf{V}}_2,
   \widetilde{\mathsf{V}}_3) \ ,
\end{equation}
where $\mathsf{Nij}$ is the Nijenhuis operator and we define
$\widetilde{\mathsf{V}} = \mathsf{G} \mathsf{V}$.

Given this definition of $\mathsf{H}$, we next define what it means to
integrate it over a 3-cycle $\Sigma$.  In the case of ordinary
three-flux this is a trivial step, since we need only to pull the
3-form $H$ back to the three-cycle and integrate it.  In the case of
the generalized flux, we will need some extra data on our three-cycle
in order to define integration.  This extra data will amount to
specifying an involutive maximal isotropic subbundle $\Omega \in T\oplus T^*$.  Roughly speaking,
$\Omega$ is needed to define the ``frame'' in which one is defining the flux.  When $\Omega = T^*$, our formulas will reduce to just the ordinary formula for $H$-flux.  When $\Omega$ includes vectors as well as forms, we will measure geometric and non-geometric fluxes.  We will call the combination of $\Sigma$ and $\Omega$ a {\em generalized
3-cycle}, $\mathsf{\Sigma}$.

Finally, we will give a prescription for integration of $\mathsf{H}$
over our generalized three-cycle $\mathsf{\Sigma}$.  As we will see,
this is the most subtle part of the story.  Because generalized
geometry is naturally covariant under T-duality, one ends up needing a
prescription for integration over a dual direction.  Such a notion of
integration can only be defined when the 3-cycle has various
isometries and we will need to put certain restrictions on the form of
the 3-cycles.  In the end, we will only give a partial
prescription for this integration, but our definition will be
sufficiently general to see that the generalized flux $\mathsf{H}$
captures all of the T-duals of $H$-flux.

Having defined a generalized notion of $H$-flux, we next present an additional
motivation for the formula (\ref{GeneralizedHFlux}).  Recalling
that $H$-flux often arises as the torsion of connection, we construct
a {\em generalized} connection $\mathsf{D}$ on $T\oplus T^*$.  This
connection is not a connection in the standard sense, since it allows
one to differentiate along T-dual directions.  Constructing what seems
to be the natural analogue of the torsion of the generalized
connection, we find that it vanishes.  However, we show that a certain
torsion-like antisymmetric object built from the connection reproduces
the generalized flux formula (\ref{GeneralizedHFlux}).

The organization of this paper is as follows: We begin with a review
of generalized geometry in section \ref{sect:ggreview}.  In section
\ref{sect:fluxdefinition} we define the generalized flux, which we
illustrate in section \ref{sect:examples} in several examples.  In
section \ref{sect:connection} we demonstrate a relationship between
the generalized flux and the generalized connection.  Finally, we
conclude in section \ref{sect:discussion} with a discussion of some
open problems and future directions.


\section{Review of generalized geometry} \label{sect:ggreview}


One motivation for introducing generalized geometry is that it is
a formalism in which T-duality acts in a simple way.  While this
formalism has been developed recently by Hitchin
\cite{Hitchin:2004ut,Hitchin:2005cv,Hitchin:2005in} and Gualtieri
\cite{Gualtieri:2003dx}, it has its roots in the older work of Duff
\cite{Duff:1989tf} and Tseytlin \cite{Tseytlin:1990nb}.  In this
section, we give an introduction to the subject which focuses on its
relationship with T-duality of the string worldsheet.  The reader
familiar with the generalized literature is warned that this
discussion is atypical and is not meant to explain the mathematical
origins of generalized geometry which are given, for example, in
\cite{Gualtieri:2003dx}.

\subsection{T-duality and generalized geometry}

We begin with a review of how T-duality acts at the level
of the classical string action.  Consider a string propagating on a
$d$-dimensional Euclidean manifold $M$, with metric $g_{\mu\nu}$ and
$B$-field $B_{\mu\nu}$;
\begin{equation} \label{eq:stringAction}
  S = \frac{1}{2} \int \, g_{\mu\nu} dX^\mu \wedge * dX^\nu +
  B_{\mu\nu} dX^\mu \wedge dX^\nu \ .
\end{equation}
When $g_{\mu\nu}$ and $B_{\mu\nu}$ do not depend on the coordinates,
$X^\mu$, we can  rewrite the action as
\begin{equation}
   S = \frac{1}{2} \int g_{\mu\nu} V^\mu \wedge * V^\nu +
  B_{\mu\nu} V^\mu \wedge V^\nu + 2 \, d\hat{X}_\mu\wedge V^\mu \ .
\end{equation}
To recover the original action (\ref{eq:stringAction}), one integrates
out $\hat{X}$, which imposes $dV = 0$.  Since every closed 1-form is
locally exact\footnote{Worrying about global issues on the worldsheet
reveals the standard exchange of winding and momentum modes and
requires that the coordinates $X^\mu$ be periodic.}, we may replace $V
= dX$, yielding (\ref{eq:stringAction}).

If instead, one integrates out $V$, one gets a new sigma model in
terms of $\hat{X}$ that is the T-dual of the original model with a new
$\hat{g}$ and $\hat{B}$ that are related to $g$ and $B$ by the Buscher
rules \cite{Buscher:1987qj,Bergshoeff:1995as}.  One also
discovers the on-shell relationship between the coordinate $X^\mu$ and its
T-dual, $\hat{X}_\mu$;
\begin{equation} \label{dXhatvsdX}
  d\hat{X}_\mu = g_{\mu\nu} *d X^\nu + B_{\mu\nu} dX^\nu \ .
\end{equation}
As noted by Duff \cite{Duff:1989tf}, if we combine $dX$ and $d\hat{X}$ into a
vector,
\begin{equation} \label{canonicalVector}
   \left(
     \begin{matrix}
       dX^\mu \\
       d\hat{X}_\mu
     \end{matrix}
   \right),
\end{equation}
then T-duality acts in a very simple way by exchanging elements of
the top with elements of the bottom.

Loosely speaking, in generalized geometry, we can define a {\em
generalized vector} to be an element $\mathsf{V} \in T\oplus T^*$,
\begin{equation}\label{generalizedVector}
   \mathsf{V}= \left(
     \begin{matrix}
       V^\mu \\
       \omega_\mu
     \end{matrix}
   \right),
\qquad V\in T, \qquad \omega \in T^* \ ,
\end{equation}
which transforms under T-duality, as well as diffeomorphisms and gauge
transformations of $B$, in the same way as (\ref{canonicalVector}).
Since it will appear often, we define $\mathsf{E} = T\oplus
T^*$\footnote{In the presence of a non-trivial $B$-field, one can only
split $\mathsf{E}$ into $ T\oplus T^*$ locally as $T^*$ is twisted by
a gerbe \cite{Hitchin:2005in}.  In the case of non-geometric spaces,
both $T$ and $T^*$ may be twisted.}.

Since $\mathsf{V}$ is a direct sum of a vector and a 1-form, we will
also write $\mathsf{V}$ as a formal sum of a vector and a 1-form,
\begin{equation}\label{VectorPlusForm}
  \mathsf{V} = V+\omega \ ,
\end{equation}
as is standard in the generalized literature
\cite{Hitchin:2004ut,Hitchin:2005cv,Hitchin:2005in,Gualtieri:2003dx}.

\subsection{Symmetries of $\mathsf{E} = T\oplus T^*$}

We now describe explicitly how various gauge transformations act on
$\mathsf{E}$. For readers familiar with the generalized literature, we
note that we are describing spacetime symmetries and not the
symmetries of the frame bundle, which can be arbitrary elements of
$O(d,d)$. 

We've already seen that, under T-duality, we just exchange forms and
vectors.  For example, if we take our spacetime to be 2-dimensional,
and T-dualize along the 1-direction, we would map
\begin{equation}\label{TdualAsMatrix}
\mathsf{V}  = \left( \begin{matrix} V^1 \\V^2 
  \\ \omega_1 \\ \omega_2 \end{matrix} \right)
\overset{\text{T}_1}{ \longrightarrow}
\left( \begin{matrix} & & 1 & \\ & 1 & & \\ 1 & & & \\ & & & 1 \end{matrix} \right) 
\left( \begin{matrix} V^1 \\V^2 \\ \omega_1 \\ \omega_2 \end{matrix} \right)
 = 
 \left( \begin{matrix} \omega_1 \\V^2 \\ V^1 \\ \omega_2 \end{matrix} \right).
\end{equation}
It is important to remember that T-duality will only be an allowed
transformation when the direction we are T-dualizing along is a $U(1)$
isometry.  As we will see later, this will require that the
generalized vectors we are dualizing be independent of the
$U(1)$-isometry direction.  Whenever we speak of an object
transforming covariantly under T-duality, we will always mean this
restricted sense.  T-dualities thus form a discreet set of global
symmetries.

We also have two kinds of local symmetries, diffeomorphisms and gauge
transformations of $B$.  Diffeomorphisms act in the natural way on the
vector and form indices.  Explicitly, if we transform coordinates from
$X^\mu$ to $X^{\mu'}$ and define $M^{\mu'}{}_\mu = \partial
X^{\mu'}/\partial X^\mu$ then $\mathsf{V}$ transforms as
\begin{equation}\label{diffasMatrix}
  \mathsf{V} \to \left(
  \begin{matrix}
    (M^{-1})^\text{t} & 0 \\
    0 & M
  \end{matrix}
\right) \,\mathsf{V} \ .
\end{equation}

Under gauge transformations $B \to B+d\lambda$,\footnote{We should
also include the large gauge transformations, $B \to B+\delta B$ which
are closed, but not exact, and are in integer cohomology.} we can see
from (\ref{dXhatvsdX}) that $d\hat{X}_\mu \to d\hat{X}_\mu +
(d\lambda)_{\mu\nu} dX^\nu$.  Thus, for the generalized vector
(\ref{generalizedVector}), we should shift $\omega_\mu \to \omega_\mu
+ (d\lambda)_{\mu\nu} V^\nu$.  This can be written in matrix form as
\begin{equation} \label{BasMatrix}
  \mathsf{V} \to 
\left(
  \begin{matrix}
    1 & 0 \\
    d\lambda & 1 
  \end{matrix}
\right)  \,
\mathsf{V} \ .
\end{equation}
Another standard notation for a gauge transformation of $B$, which
uses the notation introduced in (\ref{VectorPlusForm}) and is common
in the generalized literature, is to write
\begin{equation}
  e^{\delta B} (V+\omega) = V+\omega+i_V \delta B \ ,
\end{equation}
where $\delta B = d\lambda$, and we consider $\delta B$ to be acting from the left
by contracting indices with vectors.  Note that, as is standard, for a
form $\rho_{\mu_1 \ldots \mu_n}$, we define $(i_V \rho)_{\mu_2 \ldots
\mu_n} = V^{\mu_1} \rho_{\mu_1 \ldots \mu_n}$.

\subsection{The canonical inner product and the metric $\mathsf{G}$}

The diffeomorphisms, $B$-transformations and T-duality transformations
are all symmetries of the canonical inner product given by
\begin{equation}\label{canonicalInnerProduct}
  \langle \mathsf{V}_1 ,\mathsf{V}_2 \rangle =
  \langle V_1+\omega_1 , V_2 +\omega_2 \rangle =
\omega_1(V_2)+\omega_2(V_1) \ ,
\end{equation}
where $\omega(V) = V^\mu \omega_\mu$.  This metric has signature
$(d,d)$ and is thus invariant under local rotations in $O(d,d)$.  Note
that the full local $O(d,d)$ symmetry is only partially generated by
(\ref{TdualAsMatrix}), (\ref{diffasMatrix}) and (\ref{BasMatrix}).
The $U(1)$-isometry condition on the T-duality transformation, as well
as the requirement that the $B$-transformations are pure-gauge, put
various restrictions on the allowed symmetries.  These extra
conditions are required for the Courant bracket (to be introduced
presently) to transform covariantly.

A subbundle $\Omega \in \mathsf{E}$ on which the canonical metric vanishes is
said to be isotropic.  It is said to be maximally isotropic if its dimension is half that of $\mathsf{E}$.
Simple examples of maximally isotropic subbundles are $T^*$ and $T$.

So far, we have been ignoring the fact that $dX$ and $d\hat{X}$ are
not independent fields, and it might seem that we need to impose
(\ref{dXhatvsdX}) to project out some of the generalized vectors.  In
fact, however, the condition (\ref{dXhatvsdX}) naturally defines two
conditions $d\hat{X}_\mu = \pm g_{\mu\nu} dX^\nu+ B_{\mu\nu}dX^\nu$
depending on whether we are studying {\em right-moving} or {\em left-moving}
fields.  In terms of a generalized vector $\mathsf{V}$, of the form
(\ref{generalizedVector}), these two conditions $\omega_\mu = \pm
g_{\mu\nu} V^\nu+ B_{\mu\nu} V^\nu$ restrict $\mathsf{V}$ to be in one
of two subspaces $\mathsf{C}^{\pm} \subset \mathsf{E}$;
\begin{equation}
\mathsf{C}^{\pm} \equiv \text{span} \left\{
\left(
     \begin{matrix}
       V^\mu \\
       \pm g_{\mu\nu} V^\nu+ B_{\mu\nu}V^\nu
     \end{matrix}
   \right)
\biggl|\,\, V^\mu \in T
\right\}.
\end{equation}
Conveniently, these two spaces are orthogonal under the inner product
(\ref{canonicalInnerProduct}) and satisfy $\mathsf{C}^+\oplus
\mathsf{C}^- = \mathsf{E}$. They therefore define a {\em splitting}.

This splitting can be encoded by a matrix $\mathsf{G}$ which has
eigenvalues $+1$ for elements in $\mathsf{C}^+$ and eigenvalues of
$-1$ for elements of $\mathsf{C}^-$.  Explicitly,
\begin{equation} \label{bigG}
   \mathsf{G} = \left(
\begin{matrix}
  -g^{-1} B & g^{-1} \\
  g-Bg^{-1} B & Bg^{-1}
\end{matrix}
\right).
\end{equation}
Note that $\mathsf{G}^2 = 1$, which follows from the fact that its
eigenvalues are $\pm 1$.  Heuristically, $\mathsf{G}$ should be
thought of as the analogue of the Hodge star, $*$, on the world sheet.
In generalized geometry, we never speak of the metric and $B$-field
separately, it is only the combination (\ref{bigG}) which enters the
story.  If $\mathsf{\Lambda}$ is some combination of diffeomorphisms,
$B$-transformations and T-dualities, $\mathsf{G}$ transforms as
$\mathsf{G} \to \mathsf{\Lambda} \mathsf{G} \mathsf{\Lambda}^{-1} $.

Using $\mathsf{G}$ one can define a positive-definite inner product on
$\mathsf{E}$;
\begin{equation} \label{bigGInnerProduct}
\mathsf{G}(\mathsf{A},\mathsf{B}) = 
  \langle \mathsf{A}, \mathsf{G}\mathsf{B} \rangle
 = \langle \mathsf{G} \mathsf{A}, \mathsf{B} \rangle \ .
\end{equation}
This inner product often acts as the generalized version of the metric
$g$.

\subsection{The Courant-bracket and the Nijenhuis operator}

A basic object in generalized geometry, whose properties are discussed
in detail in \cite{Gualtieri:2003dx}, is the Courant-bracket,
\begin{equation}\label{eq:CourantBracket}
  [V_1+\omega_1,V_2+\omega_2]_C = 
        [V_1,V_2]_L + 
        \mathcal{L}_{V_1} \omega_2-\mathcal{L}_{V_2} \omega_1 
        - \tfrac{1}{2} (d(i_{V_1} \omega_2)-d(i_{V_2} \omega_1)) \ ,
\end{equation}
where $[V_1,V_2]_L$ is the Lie-bracket of two vector fields and
$\mathcal{L}_V = i_V d + di_V$ is the Lie-derivative.  
As with the
other objects we have defined, the Courant-bracket is covariant under
diffeomorphisms, $B$-transformations and T-duality.  The covariance under
diffeomorphisms is manifest from the definition, while the convariance under
gauge transformations of $B$ follows from an identity proved in \cite{Gualtieri:2003dx},
\begin{equation}\label{BtransformOfCourant}
  [e^{\delta B} \mathsf{A},e^{\delta B}\mathsf{B}]_C = e^{\delta B} [\mathsf{A},\mathsf{B}]_C
+ i_{\pi(\mathsf{B})} i_{\pi(\mathsf{A})} \delta H \ ,
\end{equation}
where $\delta H = d\delta B$ and we define $\pi:T\oplus T^* \to T$ to be the
projection onto the tangent bundle; in other words, $\pi(V+\omega) =
V$.  When $d\delta B = \delta H =0$, we find
\begin{equation}
  [e^{\delta B} \mathsf{A},e^{\delta B}\mathsf{B}]_C = e^{\delta B}
  [\mathsf{A},\mathsf{B}]_C \ , \qquad (d\delta B = 0)
\end{equation}
which is the expection result for covariance.

That the Courant-bracket is covariant under T-dualities may be
surprising since there is a theorem in \cite{Gualtieri:2003dx} which
states that gauge transformations of $B$ and diffeomorphisms are the
only allowed symmetries.  However, this theorem does not
allow for the extra assumption that there are isometries.

Indeed, suppose that we have an isometry along the $x$ direction so
that the components of our generalized vectors satisfy
\begin{equation}
  \partial_x V_i = \partial_x \omega_i = 0 \ .
\end{equation}
Let $\mu$ run over the other coordinates besides $x$ and define
 $W+\chi = [V_1 + \omega_1,V_2+\omega_2]$ .  We can then expand out
 (\ref{eq:CourantBracket}) to give
\begin{align}
   W^x &= V_1^\nu \partial_\nu V^x_2 - (1\leftrightarrow 2) \ ,
\\
   W^\mu &= V_1^\nu \partial_\nu V_2^\mu- (1\leftrightarrow 2) \ ,
\\
  \chi_x &= V_1^\nu \partial_\nu \omega_{2x}- (1\leftrightarrow 2) \ ,
\\
 \chi_\mu &= V_1^\nu \partial_\nu \omega_{2 \mu} 
  + \tfrac{1}{2} \partial_\mu (V^x \omega_x) 
  + \tfrac{1}{2} \partial_\mu (V^\mu \omega_\mu)- (1\leftrightarrow 2) \ .
\end{align}
Note that switching $V^x_i \leftrightarrow \omega_{ix}$ switches $W^x
\leftrightarrow \chi_x$ while $W^\mu$ and $\chi_\mu$ are left alone.
This yields our desired formula:
\begin{equation}
  [\mathsf{T}_x (V_1 + \omega_1),\mathsf{T}_x ( V_2 + \omega_2)]_C
 = \mathsf{T}_x[V_1 + \omega_1, V_2 + \omega_2]_C \ .
\end{equation}
We emphasize again that this formula only holds when the generalized
vectors are independent of the direction we are dualizing.

An interesting property of the Courant-bracket is that it does not
satisfy the Jacobi identity.  Rather \cite{Gualtieri:2003dx},
\begin{equation}
  [[\mathsf{V}_1,\mathsf{V}_2]_C,\mathsf{V}_3]_C +\text{cyclic}
 = d \nij(\mathsf{V}_1,\mathsf{V}_2,\mathsf{V}_3) \ ,
\end{equation}
where the Nijenhuis operator is defined by
\begin{equation} \label{nij}
\nij(  \mathsf{V}_1,\mathsf{V}_2,\mathsf{V}_3)
=
\tfrac{1}{3} \langle [\mathsf{V}_1,\mathsf{V}_2]_C,\mathsf{V}_3\rangle
+\text{cyclic} \ .
\end{equation}
The Nijenhuis operator, as we will see later, plays an important role
in defining the generalized flux.

Given a isotropic subbundle $\Omega \in \mathsf{E}$, the bundle is said to be involutive
if it is closed under the Courant bracket.  An important property of the Nijenhuis operator
is that it vanishes on an isotropic subbundle if and only if the the subbundle is
involutive \cite{Gualtieri:2003dx}.

\section{Defining the generalized NS-NS flux} \label{sect:fluxdefinition}

Computing the flux associated with the 3-form $H = dB$ can be thought
of as having three ingredients: the flux $H$, the cycle $\Sigma$ we
wish to integrate it over and the actual integration $\int_\Sigma H$.
Lifting this computation to generalized geometry requires modifying
each of these notions.

\subsection{The generalized $p$-cycle}

Let's begin by extending the notion of a $p$-cycle.  A generalized
$p$-cycle will be given by two ingredients.  The first is just an
ordinary $p$-dimensional manifold $\Sigma$ which is a submanifold of
the spacetime manifold $M$ equipped with a metric and
$B$-field\footnote{In the case of a T-fold, this definition is
inadequate since $M$ is no longer a manifold.  In the cases we will
consider, we will take $M$ to be three-dimensional and the cycle
$\Sigma$ which wraps $M$ to be just identified with $M$ itself.  We
will not attempt to give a rigorous definition here of what it means,
in general, for a T-fold to have a ``sub-T-fold''.}.

Given such a $\Sigma$, we can try to pull back the bundle $\mathsf{E}
= T_M\oplus T^*_M$ to $\Sigma$.  There is a slight subtlety in doing
this; in the presence of a nontrivial $B$-field, the bundle
$\mathsf{E}$ is twisted by a gerbe \cite{Hitchin:2005in}.  However, since we can
pull back the $B$-field to $\Sigma$, we can just put locally
$\mathsf{E}_\Sigma = T_\Sigma \oplus T^*_\Sigma$, where it is
understood that globally $T^*_\Sigma$ is twisted by the pullback of
the $B$-field.  We can also pull back the splitting of $\mathsf{E}$
into $\mathsf{C}^+ \oplus \mathsf{C}^-$.  This is accomplished by
pulling back $g$ and $B$ to $\Sigma$ and then constructing the matrix
$\mathsf{G}$ given in~(\ref{bigG}).

Given our 3-cycle, $\Sigma$ and its associated bundle,
$\mathsf{E}_\Sigma$, we define a {\em generalized} 3-cycle\footnote{This
definition of a generalized 3-cycle should be compared with
Gualtieri's different definition of a generalized (complex)
submanifold \cite{Gualtieri:2003dx}.}, by specifying a
``frame'', $\Omega$, which we take to be a maximal isotropic
subbundle $\Omega \subset \mathsf{E}_\Sigma$.  $\Omega$ is to be
thought of as the analogue of $T^*$.  The idea is that if we found
some set of T-dualities, diffeomorphisms and B-transformations which
take $\Omega$ to $T^*_{\Sigma}$ , then our definition of a
three cycle should reduce to an ordinary three cycle and the
generalized flux should reduce to the standard $H$-flux.

The reader might ask why we {\em need} to introduce $\Omega$ as an extra
piece of data.  The necessity of including $\Omega$ follows from the
fact that {\em one can construct manifolds which have multiple kinds
of flux}.  The choice of a frame $\Omega$ is just the right
extra information to select which of these fluxes we wish to measure.

To specify our choice of $\Omega$, it will be useful to introduce a
vielbein, $\mathsf{V}_i$, which spans $\Omega$ and satisfies
\begin{align}
  \mathsf{G}( \mathsf{V}_i\,, \mathsf{V}_j )
   &=\delta_{ij} \ , \qquad i\in \{1,2,3\} \ .
 \label{orthonormality}
\end{align}
Such a choice of vielbein will typically not exist globally, but we will check in
the flux-formula that we write down that we have an invariance under
$\mathsf{V}_i \to O_i{}^j\mathsf{V}_j$ for $O \in SO(p)$ so that
everything depends only on $\Omega$.

\subsection{The measure for integration}

In order to get some intuition for the role of the $\Omega$ in our definition 
of the generalized 3-cycle, consider the case when $\Omega  = T^*_{\Sigma}$.
In this case, the vielbein $\mathsf{V}_i$ which spans $\Omega$ is just a collection of forms;
\begin{equation} \label{ordinaryVs}
  \mathsf{V}_i = 
\left(
\begin{matrix}
  0\\
  \omega_i
\end{matrix}
\right), \qquad \omega_i\in T^*_\Sigma \ .
\end{equation}
The property (\ref{orthonormality}), using the explicit form
of $\mathsf{G}$ given in (\ref{bigG}), reduces to
\begin{equation}
  \omega_{i\mu} \omega_{j\nu} g^{\mu\nu} = \delta_{ij} \ .
\end{equation}
Thus, $\omega_{i \mu}$ is an ordinary vielbein.  To define a measure,
we can simply wedge the $\mathsf{V}$'s together, giving the volume form,
\begin{equation}
  \mathsf{V}_1 \wedge \mathsf{V}_2 \wedge\ldots \wedge  \mathsf{V}_p
= \omega_1 \wedge \omega_2\wedge \ldots \wedge \omega_p \ .
\end{equation}
If we have coordinates $\xi^{1,2,\ldots,p}$ on our space $\Sigma$, this
reduces to
\begin{multline} \label{ordinaryMeasure}
  \omega_{1\mu} \omega_{2 \nu} \ldots \omega_{p\rho}\, d\xi^\mu \wedge d\xi^\nu
\wedge \ldots
  \wedge d\xi^\rho
 = \det_{i\mu}(\omega_{i\mu})
     d\xi^1 \wedge d\xi^2
 \wedge \ldots \wedge d\xi^p
\\
=\sqrt{g} \,d\xi^1 \wedge d\xi^2\wedge \ldots
  \wedge d\xi^p \ ,
\end{multline}
where $g = \det(g_{\mu\nu})$.
This gives us a suitable measure for integrating a scalar.

To define a measure for a general set of $\mathsf{V}_i$, consider that
under T-duality along, say, the $\xi^1$ direction, the form $d\xi^1$
would be exchanged with the vector $\partial/\partial \xi^1$.  Thus an
integration measure,
\begin{equation}
  d\xi^1 \wedge d\xi^2 \wedge d\xi^3 \ ,
\end{equation}
would formally become the somewhat absurd looking
\begin{equation} \label{exoticMeasure}
  \frac{\partial}{\partial \xi^1} \wedge d\xi^2 \wedge d\xi^3 \ .
\end{equation}
This rather odd looking measure should be interpreted as telling
us that one of the directions we are integrating over is actually
a coordinate on the T-dual circle.  Indeed, it is convenient to
make the {\em formal} replacement,
\begin{equation} \label{dualdxiDef}
  \frac{\partial}{\partial \xi^\mu} \to d\hat{\xi}_\mu  \ ,
\end{equation}
where $\hat{\xi}_\mu$ is the coordinate T-dual to $\xi^\mu$.  We can then
write a vector field as $V^\mu \partial_\mu \to V^\mu d\hat{\xi}_\mu$.
Our measure, (\ref{exoticMeasure}), then takes the more visually appealing form,
\begin{equation}
  d \hat{\xi}_1 \wedge d\xi^2 \wedge d\xi^3 \ .
\end{equation}
Intuitively, as the notation suggests, we should integrate over
$\xi^{2,3}$ using standard integration while for the $\xi^1$ coordinate,
we should integrate over its T-dual, $\hat{\xi}_1$.  Postponing until
the next subsection the precise rules for doing this, we can write
down the formal measure:
\begin{equation} \label{generalizedMeasure}
  \mathsf{V}_1 \wedge \mathsf{V}_2 \wedge \ldots \wedge \mathsf{V}_p \ ,
\end{equation}
which reduces to the ordinary integration measure (\ref{ordinaryMeasure})
when the $\mathsf{V}_i$ take the form (\ref{ordinaryVs}).

\subsection{Integration over the generalized measure}
\label{Integration}

To understand how we should define integration over the generalized
measure, it is useful to consider the example of a generalized 1-cycle
parametrized by a coordinate $\xi^1$ with period $\Delta \xi^1$.  In
this case, our vielbein is a single vector, $\mathsf{V}$.  Suppose that we
take $\mathsf{V} = V$ where $V$ is a one component vector.  We would
like to define
\begin{equation}
  \int \mathsf{V} = \int V \ .
\end{equation}
Using (\ref{orthonormality}) implies that
\begin{equation}
  (V^1)^2 g_{11} = 1 \ ,
\end{equation}
so that we may take 
\begin{equation}
V = \frac{1}{\sqrt{g_{11}}} \frac{\partial}{\partial \xi^1} \ .
\end{equation}
Using the notation introduced in (\ref{dualdxiDef}) and assuming that
$g_{11}$ does not depend on $\xi^1$, we can write this as
\begin{equation}
  V = \sqrt{\hat{g}_{11}}  \,d\hat{\xi}_1 \ ,
\end{equation}
where $\hat{g}_{11} = g_{11}^{-1}$ is the metric of the dual circle as
found from the Buscher rules.  It is now clear how we can integrate over $V$.
We put
\begin{equation}
  \int V = \int \sqrt{\hat{g}_{11}}  \,d\hat{\xi}_1
 = \hat{L} \ ,
\end{equation}
where $\hat{L}$ is the length of the dual circle.  Noting that
$\hat{L} = L^{-1}$ where $L = \sqrt{g_{11}} \Delta \xi^1$ is the length
of the $\xi^1$ circle, we learn that
\begin{equation}\label{intOfPartial}
  \int d\hat{\xi}_1 \equiv \frac{1}{\Delta \xi^1} \ ,
\end{equation}
which is just what one would expect for the period of the dual circle.
This is the basic definition that will allow us to integrate over the
generalized measure.

Note that it was important that our integrand did not depend on the
direction we were integrating over.  It would be very interesting if
there were a natural definition of
\begin{equation}
  \int f(\xi)d\hat{\xi} = \text{?} \ ,
\end{equation}
but we suspect that, in general, no such definition exists.  Instead we
will insist that whenever we have an integral of the form,
\begin{equation} \label{generalIntegral}
  \int f(\xi^\mu) \,d\xi^1 \wedge \ldots \wedge d\xi^q \wedge
 d \hat{\xi}_{q+1}
   \wedge \ldots \wedge d\hat{\xi}_p \ ,
\end{equation}
that $f(\xi^\mu)$ only depends on $\xi^{1,2,\ldots,q}$ and that $\xi^{q+1,
\ldots,p}$ are periodically identified with period $\Delta \xi^i$.
We can then repeatedly apply formula (\ref{intOfPartial}) to yield
\begin{equation} \label{reducedIntegral}
  \left( 
    \prod_{i = q+1}^p \frac{1}{\Delta \xi^i}
  \right)
  \int f(\xi^\mu) \,d\xi^1 \wedge \ldots \wedge d\xi^q  \ . 
\end{equation}
This reduces the rather mysterious looking integral
(\ref{generalIntegral}) to an ordinary integral. Effectively, we are
{\em dimensionally reducing} along the circle directions
$\xi^{q+1,\ldots,p}$, which we can think of as being fibered over the
$\xi^{1,\ldots,q}$ directions.  After dimensional reduction, we can
then consistently integrate over the base space directions,
$\xi^{1,\ldots,q}$.

\subsection{The fiber condition} \label{U1isometry}

\begin{figure} 
\centerline{
\begin{picture}(173,166)(0,0)
\includegraphics{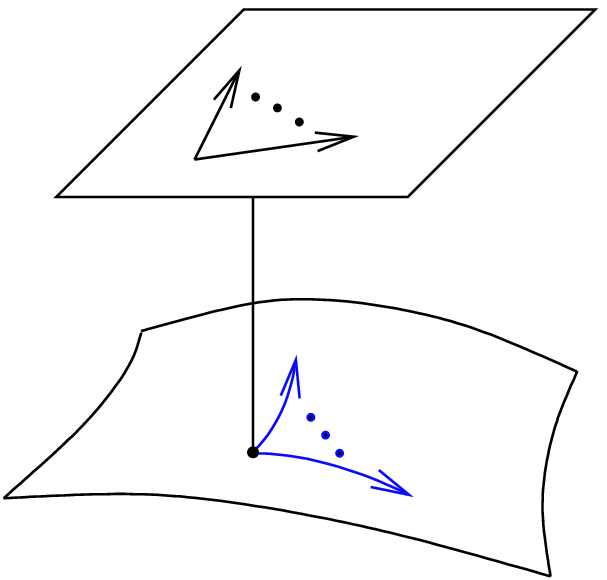}
\end{picture}
\begin{picture}(0,0)(166,0)
\put(112,20){$\xi^1$}
\put(74,70){$\xi^q$}
\put(95,127){$V_1$}
\put(58,152){$V_{p-q}$}
\put(-82,137){$T^{p-q}\text{ fiber} 
   \left\{ \text{\parbox{0in}{ \vspace{1in} }} \right.$}
\put(-90,40){$\text{Base space } 
  \left\{ \text{\parbox{0in}{ \vspace{1in} }} \right.$}
\end{picture}
}
\caption{\label{fig:fibrecondition} \footnotesize In order to define integration over our
generalized $p$-cycle, we demand that it take the form of a $T^{p-q}$
fibered over a base space. We further demand that nothing depend on
the coordinates of the torus and that the vector parts of the
generalized vielbein span the tangent bundle of the $T^{p-q}$.
Finally, we assume that the forms live in the span of the $d\xi^i$ for
$i \in \{1,\ldots,q\}$.\label{fig.U1}}
\end{figure}

To complete our definition of integration, we wish to impose that the
integral over the generalized measure can always be reduced to an
ordinary integral by repeated application of the rule
(\ref{intOfPartial}).  
In this subsection we give a simple criteria for this requirement
which we will refer to as the {\em fiber condition}.  It is quite
likely that a more general integration can be defined, but the
condition given will suffice for our purposes.

To ensure that whenever we have an integral over a dual direction, the
associated coordinate parametrizes a circle, we insist that we can
write $\Sigma$ as a $T^{p-q}$ with coordinates $\xi^{q+1, \ldots,p}$
fibered over a space with coordinates $\xi^{1,\ldots, q}$.  We then
insist that the non-zero vectors $\pi(\mathsf{V}_i)$ are a basis for
the tangent bundle of the torus fiber, while the forms live on
the base space.  Furthermore, we impose that nothing
depends on the $T^{p-q}$ fiber. These rules, which together form the
fiber condition are illustrated figure \ref{fig:fibrecondition}.

Having imposed such a strong condition on our generalized cycles, we
can ask: to what extent is the fiber condition invariant under
diffeomorphisms, $B$-transformations and T-duality?  Already with the
diffeomorphisms we see that we should restrict the diffeomorphisms to
those which preserve the torus fiber and do not depend on the torus
directions.  Generically, other diffeomorphisms will exist, but these
will take us away from the space of generalized cycles where we know
how to integrate.

When we study $B$-transformations, it is clear that we should not
allow those transformations that depend on the torus coordinates.  In
addition, recalling that the $B$-transformations take the form,
\begin{equation}
  e^{\delta B}(V+\omega) = V +\omega +i_V \delta B \ ,
\end{equation}
we see that for all $V \in T_{T^{p-q}}$ we should demand that $i_V \delta B$
is a form on the base space\footnote{Because of this restriction on
the $B$-transformations, the fiber condition is not invariant under
the expected $SO(p-q,p-q;\mathbb{Z})$ symmetry of the $T^{p-q}$ torus
fiber.  In order to restore this invariance one must define a notion
of integration that allows the dual coordinates to mix with the
coordinates in the fiber directions.}.  This ensures, for instance,
that a measure of the form,
$\prod V_i \wedge \prod \omega_j =
\prod V_i^\mu d\hat{\xi}_\mu \wedge \prod \omega_{j\mu }d\xi^\mu$,
 will be invariant under the restricted $B$-transformations, since the
shift in $V$ will be some linear combination of the $\omega_i$, which
will vanish when wedged with $\prod_j \omega_j$.

Finally, we can ask when the fiber condition is invariant under
T-duality.  First, suppose we T-dualize along one of the directions of
the fiber.  In this case, T-duality simply removes one of the
directions of the fiber and adds it to the base, taking $T^{p-q} \to
T^{p-q-1}$.  That the integral remains invariant follows by
construction from our definition of integration along the vector-like
directions.

We can also T-dualize along a base-space direction.  Suppose that the
direction we wish to T-dualize along is generated by a vector $V$.
Then, provided that $V$ doesn't depend on the fiber coordinates, it
follows that we can, at least locally, define a new coordinate
associated with the isometry, while leaving the fiber coordinates
alone.  Under T-duality, this coordinate is added to the fiber
coordinates to take $T^{p-1} \to T^{p-q+1}$.  That the integral is
unchanged follows again from our definitions.

Note that the fiber condition implies the weaker conditions,
\begin{align}
  \langle \mathsf{V}_i ,\mathsf{V}_j\rangle  &=0  \ ,
\\
  [\mathsf{V}_i,\mathsf{V}_j]_C &=0  \ .
\end{align}
These conditions are very natural since they are trivially satisfied
in the ordinary integration case when $\Omega  = T^*_{\Sigma}$.  They
are not, however, sufficient to ensure that one can perform the
generalized integral.

\subsection{The generalized version of NS-NS flux}

Having defined a generalized integral and a generalized 3-cycle, we
must now write down the flux that we wish to integrate over.  The
result, as given in the introduction, which will be motivated by the
examples, is given by
\begin{equation}
  \mathsf{H}(\mathsf{V}_1,\mathsf{V}_2,\mathsf{V}_3) = - \nij
 (\widetilde{\mathsf{V}}_1,
   \widetilde{\mathsf{V}}_2,\widetilde{\mathsf{V}}_3) \ ,
\end{equation}
where the Nijenhuis operator was defined in (\ref{nij}) and we define
\begin{equation}
  \widetilde{\mathsf{V}} = \mathsf{G} \mathsf{V} \ .
\end{equation}
The complete formula for the flux is
\begin{equation} \label{fluxFormula}
  \int_{\mathsf{\Sigma}} \mathsf{H} \equiv
  \int_\Sigma \mathsf{H}(\mathsf{V}_1,\mathsf{V}_2,\mathsf{V}_3)
  \,\, \mathsf{V}_1\wedge \mathsf{V}_2 \wedge \mathsf{V}_3 \ .
\end{equation}

Since $\Omega$ is defined to be a isotropic bundle, it follows that $\widetilde{\Omega} = \mathsf{G} {\Omega}$ is also isotropic.   Using the result of
Gualtieri \cite{Gualtieri:2003dx} that $\mathsf{Nij}$ on an isotropic subbundle is actually a tensor, 
we learn that
\begin{equation}
  \mathsf{H}(\mathsf{V}_i,\mathsf{V}_j,O_{k}{}^m\mathsf{V}_m)
   =   O_m{}^k\mathsf{H}(\mathsf{V}_i,\mathsf{V}_j,\mathsf{V}_k)
\end{equation}
for any matrix $O_i{}^j$.
Writing
\begin{equation} \label{basisindependence}
  \mathsf{H}(\mathsf{V}_1,\mathsf{V}_2,\mathsf{V}_3)
  \,\, \mathsf{V}_1\wedge \mathsf{V}_2 \wedge \mathsf{V}_3 
   = \frac{1}{3!} \sum_{i,j,k} 
   \mathsf{H}(\mathsf{V}_i,\mathsf{V}_j,\mathsf{V}_k)
  \,\, \mathsf{V}_i\wedge \mathsf{V}_j \wedge \mathsf{V}_k  ,
\end{equation}
we see that the flux is invariant under rotations $\mathsf{V}_i \to O_i{}^J \mathsf{V}_j$
provided that $O \in SO(3)$.  Hence, our flux formula only depends on $\Omega$ and
not on any particular basis\footnote{Indeed, (\ref{basisindependence}) can be defined as the projection of the
$\mathsf{Nij}$ operator onto $\widetilde{\Omega}$ using metric $\mathsf{G}(,)$.  This defines an element of $\wedge^3 \widetilde{\Omega} \in \wedge^3 \mathsf{E}$.}.

\section{Special cases of the generalized-flux formula} \label{sect:examples}

In this section we apply the flux formula (\ref{fluxFormula}) to
various specific cases in order to show that it reproduces standard
examples.  To do so, it is useful to have an explicit expression for
$\mathsf{H}$ in terms of the components of the vielbein, $\mathsf{V}_i$ .  Let our
vielbein take the form,
\begin{equation}
  \mathsf{V}_i = \left(\begin{matrix} V_i \\ \omega_i \end{matrix}\right) .
\end{equation}
We denote
\begin{equation}\label{Vtilde}
  \widetilde{\mathsf{V}}_i = \mathsf{G}\mathsf{V}_i =
   \left(\begin{matrix} \widetilde{V}_i \\ 
   \widetilde{\omega}_i \end{matrix}\right)
=
\left(\begin{matrix} g^{-1} \omega_i - g^{-1}BV_i \\ 
   gV_i - Bg^{-1}BV_i+ Bg^{-1}\omega_i \end{matrix}\right).
\end{equation}
Rather than substituting (\ref{Vtilde}) directly into the formula
for $\mathsf{H}$ and expanding it out in terms of $V_i$ and $\omega_i$
it is more useful to take the following approach: Note that
\begin{equation}
  \widetilde{\mathsf{V}}_i = 
e^B \left(\begin{matrix} \widetilde{V}_i \\ g V_i \end{matrix}\right).
\end{equation}
We can then use (\ref{BtransformOfCourant}) to find
\begin{multline} \label{expandedH}
  \mathsf{H}(\mathsf{V}_1,\mathsf{V}_2,\mathsf{V}_3) = 
\widetilde{V}_1^\mu \widetilde{V}_2^\nu \widetilde{V}_3^\rho
H_{\mu\nu\rho} +
\left[
 \widetilde{V}_1^\mu\widetilde{V}_2^\nu (V_{3\nu,\mu} - V_{3\mu,\nu})
+ \text{cyclic}
\right]
\\
-\left[\widetilde{V}_1^\mu 
  \partial_\mu(\widetilde{V}_{[2}^\nu V_{3]\nu}^{\vphantom{\nu}})
+\text{cyclic} 
\right] \ ,
\end{multline}
where in each of the terms in square brackets we must add the cyclic
permutations of $123$ and the indices are raised and lowered with $g$.
With this formula in hand, we turn to the special cases.

\subsection{Three-form flux}

The simplest case we can examine is when $\Omega = T^*$ and our vielbein takes the
form,
\begin{equation}
  \mathsf{V}_i = \left( \begin{matrix} 0 \\ \omega_i \end{matrix} \right).
\end{equation}
In this case the property (\ref{orthonormality}) reduces to
$\omega_{i\mu} \omega_{j\nu}g^{\mu\nu} = \delta_{ij}$, which implies
that the $\omega_i$ form a vielbein in the ordinary sense.  We also have
\begin{equation}
  \widetilde{\mathsf{V}}_i = \left( \begin{matrix}g^{-1} \omega_i \\
    Bg^{-1} \omega_i \end{matrix} \right) \ ,
\end{equation}
so that $\widetilde{V}_i = \omega_i^\mu$.  Since $V_i$ = 0, formula
(\ref{expandedH}) gives 
\begin{equation}
\mathsf{H}(\mathsf{V}_1,\mathsf{V}_2,\mathsf{V}_3) =
   \omega_1^\mu \omega_2^\nu \omega_3^\rho
H_{\mu\nu\rho} \ .
\end{equation}
Hence, the flux integral becomes
\begin{equation}
 \int_{\mathsf{\Sigma}} \mathsf{H}(\mathsf{V}_1,\mathsf{V}_2,\mathsf{V}_3) 
\,
  \mathsf{V}_1 \wedge \mathsf{V}_2 \wedge \mathsf{V}_3
=
 \int_\Sigma \left(\omega_1^\mu \omega_2^\nu \omega_3^\rho
H_{\mu\nu\rho} \right)
\omega_{1\gamma} \omega_{2\beta}\omega_{3\tau} \,
d\xi^\gamma \wedge d\xi^\beta \wedge d\xi^\tau \ .
\end{equation}
Noting again that the $\omega_i$ form an ordinary vielbein, this reduces to
\begin{equation} \label{Hflux}
  \int_\Sigma H \ ,
\end{equation}
which is the standard formula for 3-form flux.  Note that we did not
need to worry about the fiber condition since $V_i = 0$.

\subsection{Geometric flux}

Geometric fluxes arise from T-dualizing spaces with $H$-flux.  We
suppose that $\Sigma$ has one killing vector $V$ that generates a
circle bundle.  We then pick as our basis for $\Omega$,
\begin{equation}
  \mathsf{V}_{1,2} = \left( \begin{matrix} 0 
                             \\ \omega_{1,2} \end{matrix} \right) \ ,
\qquad
\mathsf{V}_3 = \left( \begin{matrix} V \\ BV \end{matrix} \right) \ ,
\end{equation}
where the $\omega$ are a complete set of forms on the base of the circle bundle.
One can also take $\mathsf{V}_3$ to be a pure vector; however the above choice
makes it clear that $\Omega$ is a global section of $\mathsf{E}$ when there is a
non-trivial $B$-field.  Note that we have
\begin{equation}
  \widetilde{\mathsf{V}}_3 =  \left( \begin{matrix} 0 
  \\ gV \end{matrix} \right) \ ,
\end{equation}
so that $\widetilde{V}^\mu_3 = 0$.  The vielbein property
(\ref{orthonormality}) becomes
\begin{equation} \label{VandOmegaNorm}
  V^\mu V^\nu g_{\mu\nu} = 1  \ ,
   \qquad \omega_{i\mu}\omega_{j\nu}g^{\mu\nu} = \delta_{ij} \ .
\end{equation}
It is convenient to pick one of our coordinates, $\xi^3$ to be the
circle coordinate with period 1, so that, using (\ref{VandOmegaNorm}), we have
\begin{equation}
V = \frac{1}{\sqrt{g_{33}}}\,\frac{\partial}{\partial \xi^3} \ .
\end{equation}
We can now use the formula (\ref{expandedH}) to compute the flux,
\begin{equation}
  \mathsf{H} = \omega_1^\mu \omega_2^\nu (V_{\nu,\mu} - V_{\mu,\nu})
             + \tfrac{1}{2}(\omega_1^\mu (\omega_{2} (V_3))_{,\mu} 
             + (1\leftrightarrow 2)) \ .
\end{equation}
However, using that $\Omega$ must be isotropic, we have that $\omega_{1,2} (V) = 0$, and second term vanishes.
Thus, we get just
\begin{equation}
  \mathsf{H} = \omega_1^\mu \omega_2^\nu (V_{\nu,\mu} - V_{\mu,\nu}) \ .
\end{equation}
Our measure factor $\mathsf{V}_1 \wedge \mathsf{V}_2 \wedge
\mathsf{V}_3$ gives
\begin{equation}
  \omega_{1\mu}
   \omega_{2\nu}
     B_{\rho \tau} V^\tau_3 d\xi^\mu\wedge d\xi^\nu \wedge
   d\xi^\rho+
     \frac{1}{\sqrt{g_{33}}}\,\omega_{1\mu}
   \omega_{2\nu}
       d\xi^\mu\wedge d\xi^\nu \wedge
   d\hat{\xi}_{3}  \ .
\end{equation}
The first term, however vanishes again by the isotropy of $\Omega$\footnote{{\em Proof}: We are imposing that
$\omega_{1,2}(V_3) = 0$.  We also have $\omega'_\mu = B_{\mu\nu}V^\nu$
satisfies $\omega'(V) = 0$.  However, the space of forms $\alpha$
satisfying $\alpha(V) = 0$ is only two dimensional.  Thus $\omega'$ is
a linear combination of $\omega_{1,2}$ and $\omega_1 \wedge \omega_2
\wedge \omega' = 0$.}.  Hence, we find just
\begin{equation}
   \frac{1}{\sqrt{g_{33}}} \,
\omega_{1\mu}
   \omega_{2\nu}
       d\xi^\mu\wedge d\xi^\nu \wedge d\hat{\xi}_3 \ .
\end{equation}
Putting everything together, our flux takes the form,
\begin{equation}
  \int \, 
\frac{1}{\sqrt{g_{33}}}\,\omega_1^\mu \omega_2^\nu (V_{\nu,\mu} - V_{\mu,\nu})
\,
   \omega_{1\mu}
   \omega_{2\nu}
       d\xi^\mu\wedge d\xi^\nu \wedge d\hat{\xi}_3 \ .
\end{equation}
Again using the condition that $\omega_{1,2}(V) =0$, this can be
rewritten as
\begin{equation}
   \int \,\omega_1^\mu \omega_2^\nu (dA)_{\mu\nu}
\,
   \omega_{1\mu}
   \omega_{2\nu}
       d\xi^\mu\wedge d\xi^\nu \wedge d\hat{\xi}_3 \ ,
\end{equation}
where 
\begin{equation}
  A_\mu = (\sqrt{g_{33}})^{-1} V_\mu = \frac{g_{\mu3}}{g_{33}} \ .
\end{equation}
Notice that $A_\mu$ is just the connection on the circle bundle
generated by $V$.  Examining the Buscher-rules, one notes that it is
also the T-dual of $B_{\mu3}$.

Since we have picked the length of our circle-coordinate to be one, we
may simply drop the $d\hat{\xi}_3$.  The integral then reduces to
\begin{equation}\label{geometricFlux}
  \int \,\omega_1^\mu \omega_2^\nu (dA)_{\mu\nu}
\,
   \omega_{1\mu}
   \omega_{2\nu}\,d \xi^\mu \wedge d\xi^\nu
 = \int dA \ ,
\end{equation}
where the integral is performed over the base of the circle-fibration.
This gives the first Chern-class of the circle bundle, which is the
geometric flux.

\bigskip

\noindent
{\em Example: The f-space}

\medskip
\noindent
As a simple example, consider the pure-metric space given by,
\begin{equation} \label{fMetric}
  ds^2 = (d\xi^1)^2+ (d\xi^2)^2 +
         (d\xi^3 + n \xi^1 d\xi^2)^2 \ ,
\end{equation}
The $\xi^{2,3}$ directions can be compactified in the usual
way under the symmetries $\xi^{2,3} \to \xi^{2,3}+1$.  The 
$\xi^1$ direction can also be compactified, but  under the combined
symmetry,
\begin{align} \label{xi3Identification}
  \xi^1  \to \xi^1+1 \ ,
\qquad   
  \xi^2 \to \xi^2  \ ,
\qquad  
  \xi^3 \to \xi^3 - n \xi^2 \ .  
\end{align}
This space is known variously as a Scherk-Schwarz compactification,
twisted torus and nil-manifold as well just the $f$-space
\cite{Scherk:1979zr, Kachru:2002sk, Shelton:2005cf,Grana:2006kf}.

The vielbein appropriate for measuring the geometric flux takes
a very simple form in this space;
\begin{equation}\label{fSpaceVielbeins}
   \mathsf{V}_1 = d\xi^1 \ , \qquad \mathsf{V}_2 = d\xi^2 \ , \qquad 
\mathsf{V}_3 = \frac{\partial}{\partial \xi^3} \ .
\end{equation}

Here we are treating the circle bundle as the $\xi^3$ direction.
Inspecting the metric (\ref{fMetric}), we see that $A = n\xi^1 d\xi^2$
and $dA = n d\xi^1 \wedge d\xi^2$.  Formula (\ref{geometricFlux})
reduces to just
\begin{equation}
  \int n\, d\xi^1 \wedge d\xi^2 = n \ ,
\end{equation}
which gives the geometric-flux.  It is important to note that this
integral should not be thought of as being performed over a
2-dimensional slice of the $f$-space.  In fact, no such slice exists.
The reader may check, for example that the plane determined by $\xi^3
= 0$ is not consistent with the identification
(\ref{xi3Identification}).  Rather, after we perform the ``integral''
over $\partial/\partial \xi^3$ we have effectively dimensionally
reduced along the $\xi^3$-direction so that each point specified by
$\xi^{1,2}$ corresponds to circle.

Note that T-dualizing along the $\xi^3$ direction gives a space with metric
and $B$-field,
\begin{equation}
  ds^2 = (d\xi^1)^2+ (d\xi^2)^2 +
         (d\xi^3)^2 \ ,\qquad B = n \xi^1 \,d\xi^2 \wedge d\xi^3 \ .
\end{equation}
Applying the same T-duality to the vielbein, (\ref{fSpaceVielbeins}) becomes
\begin{equation}
  \mathsf{V}_1 = d\xi^1 \ , \qquad \mathsf{V}_2 = d\xi^2 \ , \qquad 
\mathsf{V}_3 = d\xi^3 \ .
\end{equation}
Thus, to compute the generalized flux we should use (\ref{Hflux})
which gives $\int dB \,d^3\xi = n$, demonstrating the expected
invariance under T-duality.

\subsection{Non-geometric $Q$-flux}

One kind of non-geometric flux, known as $Q$-flux, which has been
studied recently
\cite{Kachru:2002sk,Shelton:2005cf,Benmachiche:2006df,Shelton:2006fd,Ellwood:2006my,Grange:2006es,Lowe:2003qy,Hull:2004in,Dabholkar:2005ve,Hull:2006va,Hull:2006qs}
is associated with a so-called $\beta$-transformation.  A
$\beta$-transformation is the double T-dual of a $B$-transformation.
It acts on generalized vectors as
\begin{equation} \label{betaTransformation}
\mathsf{V} \to e^\beta \mathsf{V} = 
  \left( \begin{matrix} 1 & \beta \\ 0 & 1 \end{matrix} \right)
\left( \begin{matrix} V \\ \omega \end{matrix} \right) ,
\end{equation}
where $\beta$ is an antisymmetric matrix.

A $Q$-space is a $T^2$ fibered over an $S^1$ in which, when one goes
around the $S^1$, one performs a $\beta$-transformation.  In order to
find global sections of $\mathsf{E}_\Sigma$, we should look for
a vielbein that is not affected by $\beta$-transformations.  Examining
the form of the $\beta$-transformation given in
(\ref{betaTransformation}), we see that generalized vectors whose
1-form part vanishes are unaffected by $\beta$-transformations.

For definiteness, let our space be a $T^2$ with coordinates
$\xi^{2,3}$ fibered over an $S^1$ with coordinate $\xi^1$.  Consider a
metric and $B$-field of the form,
\begin{equation}
  g = 
\left(
  \begin{matrix}
     1 & \\
     & g_{ab}
  \end{matrix}
\right)\ ,
\qquad
  B = 
\left(
  \begin{matrix}
     0 & \\
     & B_{ab}
  \end{matrix}
\right)\ ,
\end{equation}
where $a$, $b$ run over $2$, $3$ and nothing depends on the
coordinates $\xi^{2,3}$ of the $T^2$.  We take $\Omega$ to be
spanned by
\begin{equation}
  \mathsf{V}_1 = d\xi^1 \ , 
\qquad 
  \mathsf{V}_2 = v_1^a \frac{\partial}{\partial \xi^a} \ , 
\qquad 
  \mathsf{V}_3 = v_3^a \frac{\partial}{\partial \xi^a} \ .
\end{equation}
The property (\ref{orthonormality}) is now quite complicated:
\begin{equation}
  v_i^a (g - Bg^{-1}B )_{ab}v_j^b = \delta_{ij} \ .
\end{equation}
However, it is straightforward to find an appropriate pair of $v$'s
and substitute it into the general formula (\ref{fluxFormula}).  This
yields
\begin{equation}
  \mathsf{H}\,\, \mathsf{V}_1 \wedge \mathsf{V}_2 \wedge \mathsf{V}_3 = 
     \left[
          \frac{\partial}{\partial \xi^1} 
          \text{Re}\left(\frac{1}{\tau} \right)
     \right]
d\xi^1 \wedge 
   d\hat{\xi}_2 \wedge d\hat{\xi}_3 \ ,
\end{equation}
where we have defined $\tau = B_{12} + i \sqrt{g}$. Assuming that the
$\xi^{2,3}$ coordinates run from $0$ to $1$, we can perform the integral
over them trivially, yielding
\begin{equation} \label{Qflux}
  \text{$Q$-flux} = 
     \int \mathsf{H}\,\, \mathsf{V}_1 \wedge \mathsf{V}_2 \wedge \mathsf{V}_3
= 
\int d\xi^1
 \frac{\partial}{\partial \xi^1} \text{Re}\left(\frac{1}{\tau} \right) .
\end{equation}
To illuminate the meaning of this expression, we note that a
$\beta$-transformation acts as
\begin{equation} \label{betaOnTau}
  \tau \to \frac{\tau}{1+ \beta \tau} \ ,
\end{equation}
which takes
\begin{equation}
  \text{Re} \left(\tau^{-1}\right) \to
     \text{Re} \left(\tau^{-1}\right) + \beta \ .
\end{equation}
Since the formula (\ref{Qflux}) is an integral of a total derivative,
the $Q$-flux is given by the $\beta$-transformation that maps the top
to the bottom.  Since (\ref{betaOnTau}) must be an element of
$SL(2,\mathbb{Z})$, this gives an integer.

\bigskip

\noindent
{\em Example: The standard $Q$-space}

\medskip

\noindent
The original example of a space with $Q$-flux is found by T-dualizing
the $f$-space example (\ref{fMetric}) along the $\xi^2$-direction
\cite{Shelton:2005cf}.  This gives a metric and $B$-field,
\begin{equation}
  ds^2 = (d\xi^1)^2 + \frac{1}{1+n^2 (\xi^1)^2} ((d\xi^2)^2 + (d\xi^3)^2) \ ,
  \qquad
   B = \frac{n \xi^1}{1+n^2 (\xi^1)^2} d\xi^2 \wedge d\xi^3 \ .
\end{equation}
The $\xi^{2,3}$ directions are compactified with unit period, while
the $\xi^1$ direction is compactified with unit period only up to a
$\beta$-transformation.  The appropriate vielbein is given by
T-dualizing the vielbein in (\ref{fSpaceVielbeins}) yielding
\begin{equation}
  \mathsf{V}_1 = d\xi^1 \ , \qquad \mathsf{V}_2 = 
  \frac{\partial}{\partial \xi^2} \ , \qquad 
  \mathsf{V}_3 = \frac{\partial}{\partial \xi^3} \ .
\end{equation}
Substituting these into the flux formula yields
\begin{equation} \label{standardQflux}
  \int_{\mathsf{\Sigma}} \mathsf{H}
 = \int n\, d\xi^1 \wedge 
    d\hat{\xi}_2 \wedge
    d\hat{\xi}_3
  = n \ .
\end{equation}
To see that this agrees with the more general formula (\ref{Qflux})
note that
\begin{equation}
  \tau = \frac{n\xi^1 + i}{1+n^2 (\xi^1)^2} = \frac{1}{n\xi^1-i} \ .
\end{equation}
Hence, 
\begin{equation}
  \frac{\partial}{\partial \xi^1}\text{Re}(\tau^{-1}) = n \ ,
\end{equation}
which reproduces (\ref{standardQflux}).

\section{The generalized connection and the flux} \label{sect:connection}

In this section, we discuss a generalized connection that acts on
generalized vectors and its relation to the generalized flux.
Although the flux $H$ often arises as a torsion of a connection,
computing the analogue of the torsion of the generalized connection,
we see that it vanishes.  However, we find that the flux $\mathsf{H}$
arises from an object very similar to the torsion.

For clarity, it is useful to introduce an index notation.  We denote a
generalized vector by $\mathsf{V}^I$ where the index $I$ runs over the
tangent indices followed by the cotangent indices.  The indices can be
raised and lowered using the metric,
\begin{equation}
  \mathsf{X}_{IJ} = \left( \begin{matrix} 0 & 1 \\ 1 & 0 \end{matrix} \right),
\qquad 
  \mathsf{X}^{IJ} = \left( \begin{matrix} 0 & 1 \\ 1 & 0 \end{matrix} \right).
\end{equation}
Note that a lowered index, as in $\mathsf{V}_I$ simply runs over the
cotangent indices first followed by the tangent indices.  The matrix
$\mathsf{G}$ has index structure $\mathsf{G}^I{}_J$.  The lowered
matrix $\mathsf{G}_{IJ}$ is the positive definite metric introduces in
(\ref{bigGInnerProduct}).  The raised matrix $\mathsf{G}^{IJ}$ is,
as one would like, the inverse of $\mathsf{G}_{IJ}$ so that
$\mathsf{G}_{IJ} \mathsf{G}^{JK} = \delta_I^K$.  This follows
from the basic property that $\mathsf{G}^2 = 1$.

The goal of this section is to write down a covariant derivative
$\mathsf{D}^I$ which, when acting on vectors,
\begin{equation}
   \mathsf{D}^I \mathsf{V}^J \ , 
\end{equation}
gives a two index object covariant under diffeomorphisms,
$B$-transformations and T-duality.  Note that this is not a connection
in the ordinary sense, since it allows one to take derivatives with
respect to the T-dual coordinates.

To define the generalized connection, we begin by defining an ordinary
connection on $\mathsf{E}$.  This connection will be invariant under
diffeomorphisms and $B$-transformations, but will not be invariant
under T-duality.

We take the connection to be of the form
\begin{equation}
  D_{\mu} = \partial_\mu + \Omega_\mu \ ,
\end{equation}
where $\Omega_\mu$ is a matrix $\Omega_\mu{}^I{}_J$ which acts on the
generalized vector indices.  When the $B$-field vanishes, it
is very natural to take the connection to have the form
\begin{equation}
  D_\mu \Bigr|_{B = 0} = 
\left(
\begin{matrix} \nabla_\mu & 0 \\ 0 & \hat{\nabla}_\mu \end{matrix}
\right),
\end{equation}
where $\nabla_\mu$ is the Levi-Civita connection on vectors and
$\hat{\nabla}_\mu$ is the Levi-Civita on 1-forms.

When $B \ne 0$, one can partially fix the form of $D_\mu$ by demanding
that it annihilate both $\mathsf{X}^{IJ}$ and $\mathsf{G}^{IJ}$ and
that it transform covariantly under $B$-transformations.  This
unfortunately is not enough to completely determine the connection, as
one is still left with a one-parameter family of possible connections:
\begin{equation} \label{connectionFamily}
\left(
  \begin{matrix}
     1 & 0 \\ B & 1
  \end{matrix}
\right)
\left[
\left(
\begin{matrix} \nabla_\mu & 0 \\ 0 & \hat{\nabla}_\mu \end{matrix}
\right)
+a
\left(
\begin{matrix} 0 & \frac{1}{2}g^{-1} H_\mu g^{-1} \\ \frac{1}{2} H_\mu & 0 \end{matrix}
\right)
\right]
\left(
  \begin{matrix}
     1 & 0 \\ -B & 1
  \end{matrix}
\right).
\end{equation}
Here we have used the shorthand $H_\mu$ for $H_{\mu\nu\rho}$ where
$\nu$ and $\rho$ are treated as matrix indices.  To fix an appropriate
choice for $a$, it is useful to turn to string theory for guidance.
Recall that in the fermionic terms of the $N = 1$ string action the
kinetic terms use the connection
\begin{equation}
  \nabla^{\pm}_\mu = \nabla_\mu \pm \tfrac{1}{2} g^{-1} H_\mu \ ,
\end{equation}
where we take $+$ for the right moving fermions and $-$ for the left
moving fermions.  This connection, known as the {\em Bismut
connection} in the generalized literature, was first introduced in
string theory by Gates, Hull and Rocek \cite{Gates:1984nk} and is
relevant for a number of applications in generalized complex geometry
\cite{Gualtieri:2003dx, Hitchin:2005in}.

We can now fix the form of (\ref{connectionFamily}) by insisting that 
if $\mathsf{V} \in \mathsf{C}^{\pm}$ that
\begin{equation}
  \pi(D_\mu \mathsf{V}) =  \nabla^{\pm}_\mu \pi(\mathsf{V}) \ .
\end{equation}
In other words, the covariant derivative just acts as the Bismut
connection on the vector part of $\mathsf{V}$.  This extra condition
fixes $a = 1$ and gives the lift of the Bismut connection to
generalized geometry:
\begin{equation} \label{generalizedBismut}
D_\mu \equiv
  \left(
  \begin{matrix}
     1 & 0 \\ B & 1
  \end{matrix}
\right)
\left(
\begin{matrix} \nabla_\mu & \frac{1}{2}g^{-1} H_\mu g^{-1} 
  \\ \frac{1}{2} H_\mu & \hat{\nabla}_\mu \end{matrix}
\right)
\left(
  \begin{matrix}
     1 & 0 \\ -B & 1
  \end{matrix}
\right).
\end{equation}
This connection has nice properties under T-duality.  Suppose the
$x$-direction parametrizes a circle and that neither $g$ nor $B$
depends on $x$.  Then we find\footnote{We have not found a simple
proof of this formula.  However, it is straightforward, although
cumbersome, to check by a direct application of the Buscher-rules.}
\begin{align}
\label{dualOfDmu}
  D^\mu 
    &\overset{\text{T}_x}{\longrightarrow} 
       \mathsf{T}_x D^\mu  \mathsf{T}_x \ ,
\qquad
  \mu\ne x \ ,
\\
\label{dualOfDx}
  D^x &\overset{\text{T}_x}{\longrightarrow}
\mathsf{T}_x \left( 
B_{x\sigma} D^\sigma - \mathsf{G} g_{x\sigma}D^\sigma
\right)\mathsf{T}_x  \ .
\end{align}
The matrices $\mathsf{T}_x$ are the elements of $SO(d,d)$ which
represent T-duality along the $x$ direction.  Since T-duality switches
vectors with forms, (\ref{dualOfDx}) gives us the form part of the
generalized connection.  Indeed, setting
\begin{equation}\label{generalizedConnection}
  \mathsf{D}^I = \left( \begin{matrix} 
D^\mu \\
 - \mathsf{G} D_\mu+ B_{\mu\sigma} D^\sigma 
\end{matrix}
\right),
\end{equation}
it is straightforward to check using (\ref{dualOfDmu}) and
(\ref{dualOfDx}) that, when acting on a vector, as in $\mathsf{D}^I
\mathsf{V}^J$, that the resulting two index object transforms
covariantly under diffeomorphisms, $B$ transformations and T-duality.

\subsection{Parallel transport and torsion}

We define the parallel transport of $\mathsf{V}_2$ along
$\mathsf{V}_1$ by
\begin{equation} \label{parallelTransport}
  \mathsf{D}_{\mathsf{V}_1} \mathsf{V}_2^K = \mathsf{X}_{IJ} 
  \mathsf{V}^I_1 \,\mathsf{D}^J \mathsf{V}^K_2 \ .
\end{equation}
This definition of parallel transport has a nice formula in terms of
the connection (\ref{generalizedBismut}), which can be found using the
definitions (\ref{generalizedConnection}) and
(\ref{parallelTransport}):
\begin{equation}
  \mathsf{D}_{\mathsf{V}_1} \mathsf{V}_2^K
 = \pi(\widetilde{\mathsf{V}}_1)^\mu D_\mu \mathsf{V}^K_2
 - \pi(\mathsf{V}_1)^\mu D_\mu \widetilde{\mathsf{V}}^K_2 \ .
\end{equation}
This expression gives a derivation of the ``Leibniz rule'',
\begin{equation}
  \mathsf{D}_{\mathsf{V}_1} (f \mathsf{V}_2)
 = f \mathsf{D}_{\mathsf{V}_1}  \mathsf{V}_2
 + [\pi(\widetilde{\mathsf{V}}_1)(f)] \mathsf{V}_2 - [\pi(\mathsf{V}_1)(f) ]
\widetilde{\mathsf{V}}_2 \ .
\end{equation}

Now that we have defined parallel transport, we may attempt to 
define a torsion
\begin{equation} \label{torsionAnsatz}
 \mathsf{T}(\mathsf{V}_1, \mathsf{V}_2) = \mathsf{D}_{\mathsf{V}_1}
 \mathsf{V}_2 - \mathsf{D}_{\mathsf{V}_2} \mathsf{V}_1
 - [\mathsf{V}_1 ,\mathsf{V}_2] \ .
\end{equation} 
A nice choice for the bracket, $[,]$, which makes $\mathsf{T}$ into a
tensor, is given by
\begin{equation} \label{newBracket}
  [\mathsf{V}_1 ,\mathsf{V}_2 ] = 
\mathsf{G}[\widetilde{\mathsf{V}}_1,\widetilde{\mathsf{V}}_2]_C -
\mathsf{G}[\mathsf{V}_1,\mathsf{V}_2]_C  \ . 
\end{equation}
A straightforward, but tedious computation of $\mathsf{T}(\mathsf{V}_1,
\mathsf{V}_2)$ reveals that
\begin{equation}
   \mathsf{T}(\mathsf{V}_1,
\mathsf{V}_2) =0 \ ,
\end{equation}
so that, in the sense defined by (\ref{torsionAnsatz}) and
(\ref{newBracket}), the torsion vanishes.

This computation suggests that the notion that the flux is given by
the torsion of the connection, as holds for the Bismut connection for
example, is not quite right.  Consider, however, the torsion-like
quantity\footnote{For the Bismut connection, for example, we would find
$\nabla^\pm_{V_1} V_2^\mu V_{3\mu} - \nabla^\pm_{V_2} V_1^\mu V_{3\mu}
 = \pm H(V_1,V_2,V_3) + [V_1,V_2]^\mu V_{3\mu}$, yielding the torsion plus
a term that vanishes provided $[V_1,V_2] = 0$.},
\begin{equation}
  -\tfrac{1}{3}\left(\langle  (\mathsf{D}_{\mathsf{V}_1} \mathsf{V}_2),
   \mathsf{V}_3 \rangle
-\langle  (\mathsf{D}_{\mathsf{V}_2} \mathsf{V}_1),
\mathsf{V}_3 \rangle
\right) 
+ \text{cyclic} \ .
\end{equation}
Using (\ref{torsionAnsatz}) and (\ref{newBracket}) this reduces to
\begin{equation} \label{torsionLikeThing}
  \mathsf{H}(\mathsf{V}_1,
  \mathsf{V}_2,\mathsf{V}_3)
-\tfrac{1}{3} [\langle [\mathsf{V}_1,\mathsf{V}_2]_C,\widetilde{V}_3 \rangle
+\text{cyclic} ] \ .
\end{equation}
Notice that for $\mathsf{V}$'s which are appropriate for a generalized
3-cycle, we would have $[\mathsf{V}_i,\mathsf{V}_j]_C = 0$, so that
(\ref{torsionLikeThing}) would reduce to just
$\mathsf{H}(\mathsf{V}_1, \mathsf{V}_2,\mathsf{V}_3)$.  This is, in
fact, how we originally found the flux formula.

\subsection{Differentiation of tensors}

Although they are not relevant for the main line of discussion, we end
this section with a few observations about the action of generalized
connection on tensors. For $\mathsf{A}$ a generalized vector, we have
the following identity,
\begin{equation} \label{GonD}
  \mathsf{G}^I{}_J \mathsf{D}^J \mathsf{A}^K +
  \mathsf{D}^I (\mathsf{G}^K{}_L \mathsf{A}^L) = 0 \ .
\end{equation}
This implies that the index on the generalized connection lives in the
opposite half of the splitting as the index of the vector it is
differentiating.

Because of the $\mathsf{G}$ in the definition
(\ref{generalizedConnection}), $\mathsf{D}^I$ does not satisfy the
Leibniz rule when acting on products of vectors unless all of the
vectors live in $\mathsf{C}^+$ or all live in $\mathsf{C}^-$ (in which
case one can replace $\mathsf{G}$ by $\pm 1$).  This
implies that it is not meaningful to speak of differentiating a tensor
$\mathsf{T}^{I_1 I_2\ldots I_n}$ unless it satisfies
\begin{equation} \label{differentiableTensor}
\forall i,j \qquad
  \mathsf{G}^{I_i}{}_{J} \mathsf{T}^{I_1 \ldots I_{i-1} J I_{i+1}\ldots I_{n}}
 = \mathsf{G}^{I_j}{}_{J} 
\mathsf{T}^{I_1 \ldots I_{j-1} J I_{j+1}\ldots I_{n}} \ .
\end{equation}
Because of the rule (\ref{GonD}) we cannot act multiple times with the
connection since the property (\ref{differentiableTensor}) is not
preserved under differentiation.  This makes it very difficult to
construct a curvature of the generalized connection.

\section{Discussion} \label{sect:discussion}
We conclude with a few comments on future directions and problems that
we believe deserve further study.
\begin{enumerate}
\item In our construction of the generalized flux integral, we relied
heavily on what we called fiber condition.  This condition was
required to ensure that we could give a sensible definition of
integration over a generalized 3-cycle.  It seems likely, however,
that the spaces on which integration is well-defined could be
extended.  Currently, for example, our definition is not broad enough
to handle spaces where coordinates and dual coordinates mix on the
torus fibers and we are, thus, not able to realize a full
$SO(d,d;\mathbb{Z})$ symmetry for our definition of integration.  This
makes it difficult to understand fluxes on spaces for which there is
no geometric dual.
\item Although in the examples we were able to show that the integral
of $\mathsf{H}$ over a generalized 3-cycle was always a
topological quantity and in fact an integer, it would be nice to have
proof of this in the framework of generalized geometry.
\item Our discussion of the generalized connection seems far from
complete.  There is already a well-established connection which acts
on pure spinors \cite{Gualtieri:2003dx}, and it would be interesting
to try to connect the two.  It also seems quite interesting to try to
understand whether there is a natural notion of the curvature of the
connection.
\item It would be nice to give a stringy derivation of the flux
formula.  The string action can already be written in a generalized
form, following
\cite{Tseytlin:1990nb,Hull:2006tp,Hull:2006qs,Hull:2006va}, and the
related works
\cite{Lindstrom:2004cd,Lindstrom:2004hi,
 Bredthauer:2005zx,Merrell:2006py,Lindstrom:2006ee},
\begin{equation}
  S = \frac{1}{4} \int  \mathsf{G}_{IJ} \mathsf{Z}^I 
 \wedge * \mathsf{Z}^J + \mathsf{B}_{IJ} \mathsf{Z}^I \wedge \mathsf{Z}^J \ ,
\end{equation}
where $\mathsf{B}_{IJ}$ is the canonical anti-bracket of $\mathsf{E}$ given by
the matrix,
\[
\mathsf{B}_{IJ} = \left(\begin{matrix} 0 & 1 \\ -1 & 0 \end{matrix} \right),
\]
and $\mathsf{Z} = dX^\mu + \Omega_\mu$ where $\Omega_\mu$ is an
auxiliary one-form on the worldsheet as well as in spacetime.  It
would be quite nice if we could replace the $\mathsf{B}$ term with a
WZW-term involving $\mathsf{H}$, but we have not yet found a way to do
so.
\end{enumerate}

\section*{Acknowledgments}

We would like to thank Thomas Grimm, Akikazu Hashimoto, Albrecht
Klemm, Leopoldo Pando Zayas, Jessie Shelton, Bernardo Uribe and Diana
Vaman for useful conversations.  This work was supported in part by
DOE grant DE-FG02-95-ER40896 and funds from the University of
Wisconsin-Madison.

\bibliography{n}\bibliographystyle{utphys}

\end{document}